# Reconstructing fluid dynamics with micro-finite elements

Wennan ZOU

Institute for Advanced Study, Nanchang University, Nanchang, P.R. China, 330031

Email: zouwn@ncu.edu.cn

**ABSTRACT**: In the theory of Navier-Stokes equations, an incompressible fluid in motion is modeled as a homogeneous and dense assemblage of constituent "fluid particles" with viscous stress proportional to rate of strain. In this paper, by virtue of the alternative constituent "micro-finite element", we introduce a set of new intrinsic quantities, called the vortex fields, to characterize the relative orientation between elements and the feature of micro-eddies on the element, while the description of viscous interaction in fluid returns to the original intuition that the interlayer friction is proportional to the slip strength. Such a framework enables us to reconstruct the dynamics theory of viscous fluid, in which the fluid can be modeled as a finite covering of elements and consequently indicated by a differential manifold that admits complex topological evolution.

**Key Words**: Micro-finite element; Vortex fields; Viscous interaction; Fluid dynamics; Topological evolution

## 1. Introduction

The fundamental hypothesis underlying fluid dynamics is that the matter of which the fluid is made up is continuously distributed and that the field variables involved, such as velocity, pressure, mass density, etc., are continuous functions of space and time. Some people believe that such a macroscopic theory of fluids is unique, as what the Navier-Stokes (N-S) equations stand for [1][2]. Previously, we recast the theory of elasticity with the so-called micro-finite elements [3] that have the property of intrinsic stretch and may finitely cover the body, implying the non-uniqueness of macroscopic models of elastic media, and the equivalence of two kinds of theories of elasticity when some reasonable compatibility conditions have been applied. But when a theory of fluid dynamics is elaborated with micro-finite elements in a similar way, namely representing the fluid in motion as a four-dimensional space-time manifold, we find that (1) the possibility to degenerate the new theory to the classical one is no longer exists; (2) the viscous interaction in fluid could be reckoned as an internal friction instead of an analogy of elastic stress. Thus, we obtain a completely different macroscopic flow theory of micro-finite fluid elements with viscous friction [4] proportional to slip strength, whilst the N-S equations only represent a flow theory of infinitesimal fluid particle with viscous stress proportional to rate of strain.

In this paper, all quantities and equations are established under the Galilean space-time $E_4 = \mathbb{R}^3 \times \mathbb{R} = \{x_1, x_2, x_3, x_4 = t | x_\mu \in \mathbb{R}, \mu = 1,2,3,4\}$ where $(x_1, x_2, x_3)$ is a Cartesian coordinate system and $x_4 = t$ stands for the time. The bases $(\mathbf{e}_1, \mathbf{e}_2, \mathbf{e}_3)$ of tangent space $T(E_3)$ are referred to as $\mathbf{e}_i = \partial\mathbf{r}/\partial x_i$ from the position vector $\mathbf{r}$. The natural bases for the vector space $\Lambda^1(E_4)$ of all 1-forms are given by $\{dx_1, dx_2, dx_3, dx_4 = dt\}$. A vector/form is denoted by a minuscule Latin bold letter or by its components, for instance $\mathbf{r}$ or $r_i$. Summation over repeated indices is tacit, from 1 to 3 for Roman indices and 1 to 4 for Greek indices. The generated bases, called the area element and volume element, for the vector spaces $\Lambda^2(E_3)$ and $\Lambda^3(E_3)$ of spatial forms are expressed by

$$da_i = \frac{1}{2}\epsilon_{ijk}\, dx_j \wedge dx_k, dv = dx_1 \wedge dx_2 \wedge dx_3, \tag{1}$$

respectively, where $\epsilon_{ijk}$ is the permutation symbol. The superscript index is used for the axial vector, such as angular velocity, moment of force, etc.

In the classical fluid dynamics, it is confirmed that the transformation of an addition of constant velocity onto the fluid does change the acceleration and the viscous stress determined by the velocity gradient, and so has no effect on the dynamical process of the situation [5]. But as Fig. 1 shown, the contact relation or slip state in the fluid described by the streamlines become quite different for two flows after an inertial transformation. According to the new theory, the viscous interaction between the interlayers of the fluid, if measurable, will recognize a unique investigator seeing the truthful flow velocity, which is called the *contact velocity* $v_i$. In the following derivation, it is more profitable to introduce the micro-displacement 1-form $\mathbf{u}_i = v_i dt$ as a quantity.

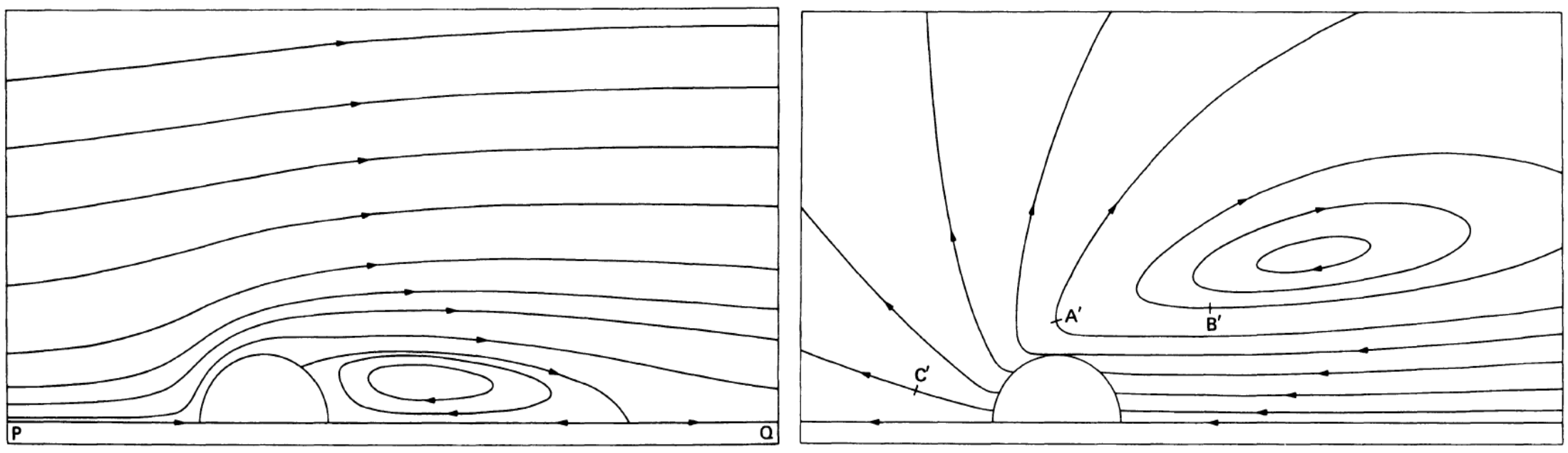

Fig. 1. The contact relations in the fluid become greatly different after an inertial transformation: (left) Flow past a circular cylinder at Re = 40, (right) Streamlines for cylinder moving through stationary ambient fluid. (Tritton, 1988, page 76)

## 2. Flow fields and their intensities

When the fluid under consideration is incompressible and viscous, as water or air flowing in everyday life, the elements, defined everywhere to describe the inhomogeneous change in the fluid, should be the same by analogy with the elasticity. But viscous slip-flow could be quite different from elastic deformation, no micro-finite element with internal layers in molecule scale and slipping over each other can be traced and identified globally; and laminarization of fluid is a kind of statistical tension making a corresponding rotation of fluid particulates when they diffuse in different directions, that is not like the elastic tension originating from the density difference of atoms/molecules in different directions. The fluid in motion is then the result of pasting together micro-finite elements confined to local space-time [6]. The velocity $v_i$, now uniquely identified as the contact velocity, is just the extrinsic property of flowing elements, instead of the displacement. Besides, once the shear-induced order [7][8] has been formed and/or the micro-eddies are wheeling, the vortex fields are naturally introduced to characterize the relative orientation between neighboring elements and the micro-rotation in the element, which are regarded as the structural and intrinsic properties of elements in flow, as shown in Fig. 2.

The vortex fields are indicated by an axial-vector valued differential 1-form

$$\mathbf{W}^i = W_\mu^i dx_\mu = \Phi^i dt + A_k^i dx_k, \tag{2}$$

where the temporal part $\Phi^i$ is referred to as the micro-eddy field while the spatial part $A_k^i$ the swirl field. Mathematically, the vortex fields correspond to the connection on frame bundle of fluid manifold, which is

$$D\mathbf{e}_i = \epsilon_{ijk}\mathbf{W}^j\mathbf{e}_k, \tag{3}$$

where $D$ is the covariant exterior differential operator. Using the swirl field, one can calculate the slip strength by

$$D(u_i\mathbf{e}_i) = (Dv_i)\mathbf{e}_i \wedge dt = Y_{ki}\mathbf{e}_i dx_k \wedge dt, \tag{4}$$

where '∧' stands for the exterior product, and

$$Y_{ki} = D_k v_i = \partial_k v_i + \epsilon_{ilm} A_k^l v_m \tag{5}$$

is referred to as the velocity intensity. For the Newtonian (linear) fluids, the viscous friction is formulated as

$$\boldsymbol{\sigma} = \mu * D(\mathbf{u}_i\mathbf{e}_i) = \mu(D_k v_i)\mathbf{e}_i da_k = \mu\big(\partial_k v_i + \epsilon_{ilm} A_k^l v_m\big)\mathbf{e}_i da_k, \tag{6}$$

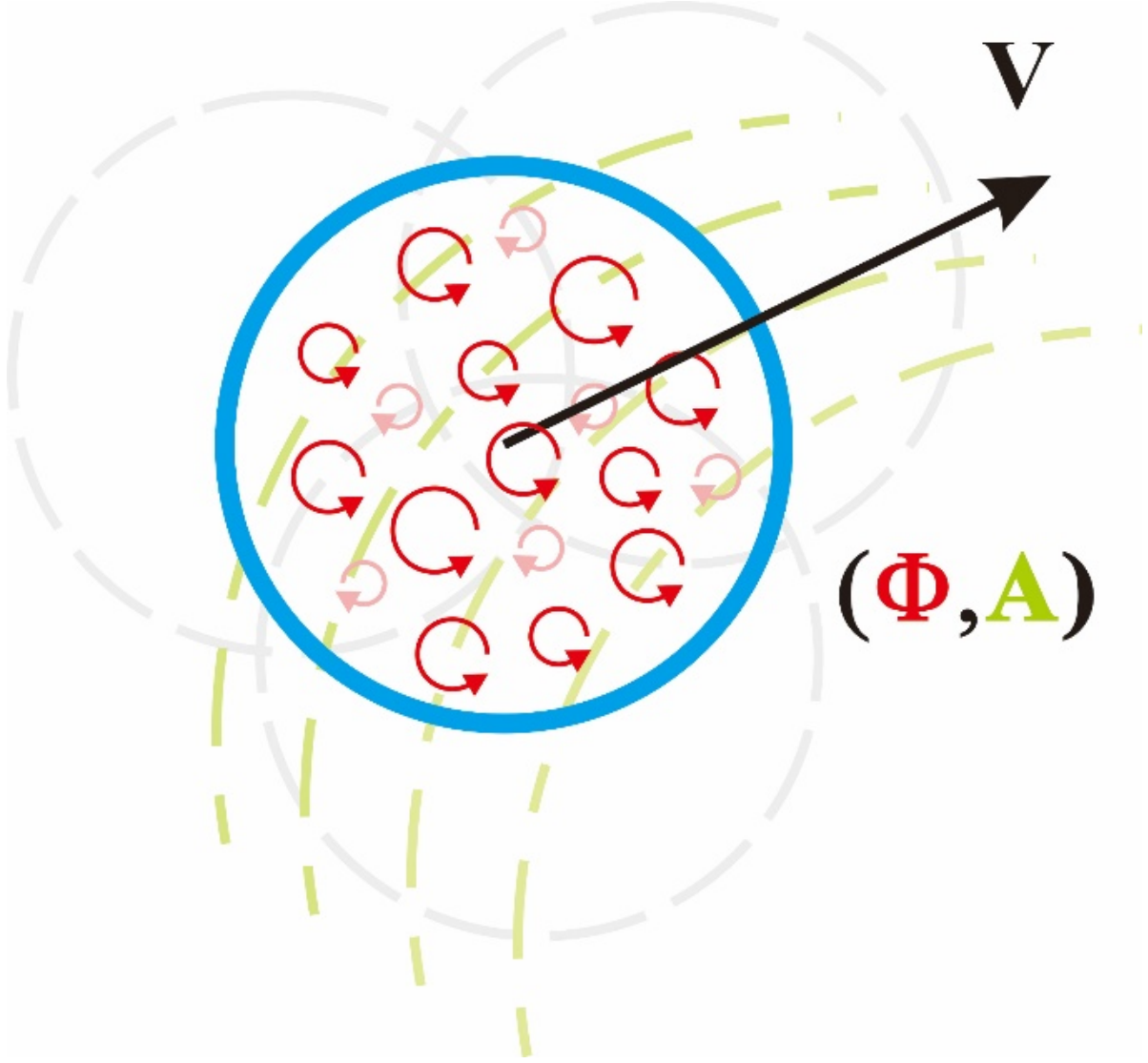

Fig. 2. Sketch of the micro-finite element.

where the operator '$*$' for any base form $b$ is defined by

$$b \wedge * b = dv \wedge dt. \tag{7}$$

The topological structures of the connection are essentially determined by the curvature tensor, namely the vortex intensity defined by

$$\mathbf{F}^i = D\mathbf{W}^i = d\mathbf{W}^i + \epsilon_{ijk}\mathbf{W}^j \wedge \mathbf{W}^k = B_k^i da_k + H_k^i dx_k \wedge dt, \tag{8}$$

With the spatial part $B_k^i = \epsilon_{klm}\left(\partial_l A_m^i + \frac{1}{2}\epsilon_{ipq}A_l^p A_m^q\right)$ referred to as the swirl intensity and the spatio-temporal part $H_k^i = \partial_k \Phi^i - \partial_t A_k^i + \epsilon_{ipq}A_k^p \Phi^q$ the micro-eddy intensity. From (8), the structural equations (or called the Bianchi identity mathematically) can be derived as

$$D\mathbf{F}^i = d\mathbf{F}^i + \epsilon_{ijk}\mathbf{W}^j \wedge \mathbf{F}^k \equiv 0. \tag{9}$$

Geometrically, the vortex fields consist of twelve components but an arbitrary rotation transformation of frame with three parameters is permitted (cf. Section 4), that is to say, only nine variables in the vortex fields are objective for the description of structures on the manifolds, while the derived curvature tensor is objective whose eighteen components need nine identities as given by (9) to restrict (Note: the equation (9) consists of twelve relations, but not all of them are independent).

In summary, the flow fields and their intensities can be explained as the properties on an element or between elements as

- $v_i$: the micro-displacement of element after unit time;
- $Y_{ki} = D_k v_i = \partial_k v_i + \epsilon_{ilm}A_k^l v_m$: the difference of micro-displacement of the element after unit time when translating unit distance along the direction $\mathbf{e}_k$, characterizing the nonuniformity of flow;
- $\Phi^i$: the net micro-rotation in the element due to small-scale motion after unit time;
- $A_k^i$: the required micro-rotation of the element due to the layered tension when translating unit distance along the direction $\mathbf{e}_k$;
- $B_k^i$: the micro-rotation jumping of the element when translating along a loop with unit area perpendicular to $\mathbf{e}_k$;
- $H_k^i$: the difference of left micro-rotation in the element after unit time when unit distance along the direction $\mathbf{e}_k$.

From another viewpoint, the flow fields indicate the average (statistical) state of the element with a large number of particulates, and must be attached to particulates during their microscopic diffusion, which further results in the coupling of the velocity and the vortex fields. As mentioned above, a particulate with velocity $v_i$ diffusing unit distance along the direction $\mathbf{e}_k$ (or across the unit area perpendicular to $\mathbf{e}_k$) will transport a momentum $\partial_k v_i$ plus an increment $\epsilon_{ilm}A_k^l v_m$ induced by the ordering atmosphere. The processes of a group of particulates yields the expression (6) of slip friction. Mathematically, it is simply a replacement of the ordinal differential by the covariant differential considering the connection. A further and marvelous contribution of microscopic diffusion

$$\boldsymbol{\chi}_i = DD\mathbf{u}_i = \epsilon_{ilm}B_k^l v_m da_k \wedge dt \tag{10}$$

indicates the induced jump of momentum $\chi_{ki} = \epsilon_{ilm}B_k^l v_m$ emerging from the ordering structure when a particulate diffusing along a loop with unit area perpendicular to $\mathbf{e}_k$. Besides, the micro-eddy indicated by $\Phi^i$ is rather different from the global translation represented by $v_i$. The former may be understood as the disorder during the ordering adjustment of particulates, though it actually opens a new way to intervene the main stream. Thus, the micro-eddy intensity $H_k^i$ can be understood as the flux of micro-rotation per time carried by the particulate diffusing unit distance along the direction $\mathbf{e}_k$.

# 3. Viscous interactions under layered tension

The vortex fields, indicating the shear-induced ordering (Fig. 3), strongly interact with the microscopic diffusion of particulates, such that some complicated structures and mechanism can be involved in the turbulence. According to the simplified model of the new theory, the global rigid rotation of a bulk of fluid, say the micro-finite element, cannot come true through the viscosity.

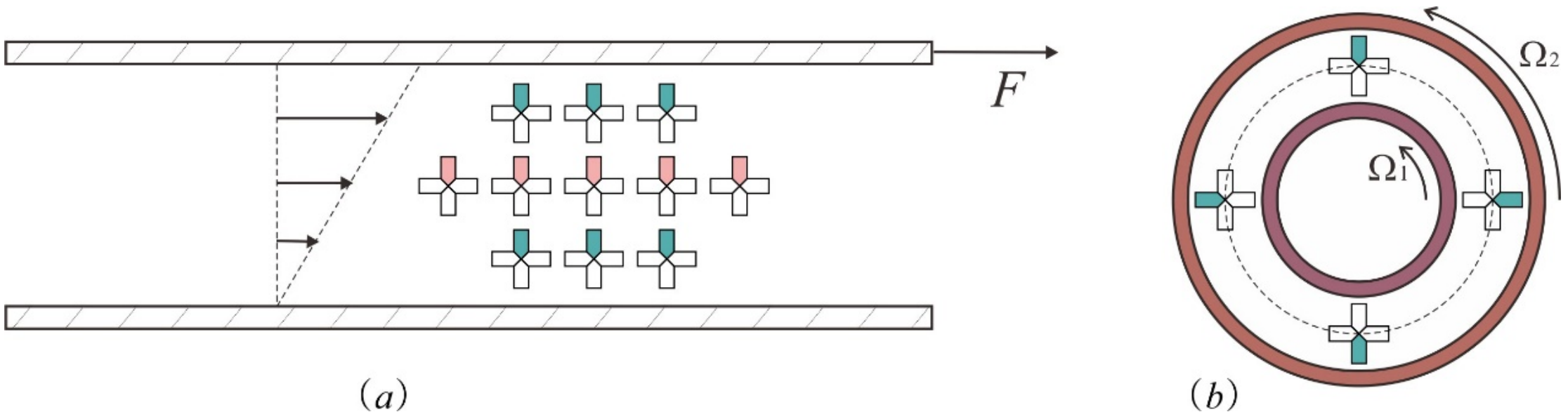

Fig. 3. Shear-induced laminarization.

The viscous interaction working under the layered tension has become diverse. Besides those activated by the microscopic diffusion of particulates, the twirling process to bind the main stream and slip friction is indicated by $J_k^i = \epsilon_{ilm}\sigma_{kl}v_m$, which makes sense when the direction of shear deviates from the direction of flow. We believe that the twirling mechanism coming from the non-parallelism of shear and flow is the origin of the complexity of turbulence. The twirling interaction is usually indicated by an axial vector 3-form $J_k^i da_k \wedge dt$ with the measure of moment of force.

From the above discussion, we can conclude that the swirl field joins the viscous interaction by coupling with the velocity while the eddy field forms an independent kind of viscous interaction. As the general routine to construct a physical theory, the variational principle from the Lagrangian density of the fluid is expectable. The merits of a variational derived theory include a certain internal consistency and that the physical symmetries of the system can be directly related to conservation laws that are satisfied by any and all solutions, last but not least the fact that variational formulations lead to a complete theory of compatible boundary and initial data. For instance, the Lagrangian density

$$\mathcal{L}_{Euler}[v_i] = v_i(\rho a_i - f_i + \partial_i p) \tag{11}$$

can be used to obtain the Euler equations

$$M_i = \frac{\partial \mathcal{L}_{Euler}}{\partial v_i} = \rho a_i - f_i + \partial_i p = 0, \tag{12}$$

for perfect and incompressible fluids, where the acceleration $a_i$, body force $f_i$ and pressure $p$ should be regarded invariant in variational process. For real fluids with viscosity, an additional Lagrangian density $\mathcal{L}_{viscosity}$ can be constructed as

$$\mathcal{L}_{viscosity}\left[v_i; W_\mu^i\right] = \mathcal{L}_Y[Y_{ki}] + \mathcal{L}_\chi[\chi_{ki}] - \mathcal{L}_\Phi\left[\Phi^i\right] - \mathcal{L}_H\left[H_k^i\right] \tag{13}$$

following the minimal replacement and the minimal coupling principles of the gauge field theory (cf. [9]), where $\mathcal{L}_Y, \mathcal{L}_\chi, \mathcal{L}_\Phi, \mathcal{L}_H$ are non-negative energy functionals of generalized fluxes $Y_{ki}, \chi_{ki}, \Phi^i, H_k^i$. In the above formulation, the micro-eddy field represents a motion with micro-scale and essentially dissipative, which is independent to the main stream. The minus signs in front of $\mathcal{L}_\Phi$ and $\mathcal{L}_H$ proclaim the transferability of the action quantities between the micro-eddies and the main stream, through the viscous interactions and with the swirl structures as a bridge. Based on the Lagrangian density

$$\mathcal{L}\left[v_i; W_\mu^i\right] = \mathcal{L}_{Euler}[v_i] + \mathcal{L}_{viscosity}\left[v_i; W_\mu^j\right], \tag{14}$$

we list the work conjugate pairs of generalized fluxes and forces in Table 1, where some relations between the generalized forces and the generalized fluxes can be regarded as constitutive laws while the others are just derived through the coupling processes, say $\Sigma_i = \frac{\partial \mathcal{L}_\chi}{\partial v_i} = \epsilon_{ilm} B_k^m \frac{\partial \mathcal{L}_\chi}{\partial \chi_{kl}}$. Comparing with the generalized flux, the associated generalized force has the same direction property as an ordinary or axial vector, but their form bases are complementary to each other, for example, $\mathbf{Y}_i \wedge \boldsymbol{\sigma}_i = Y_{ki}\sigma_{ki} dv \wedge dt$.

Table 1. Work conjugate pairs of generalized fluxes and forces in viscous interactions

| Generalized fluxes | Generalized forces | Properties |
|---|---|---|
| $v_i$ | $\Sigma_i = \partial \mathcal{L}_\chi / \partial v_i = \epsilon_{ilm} \Pi_{kl} B_k^m$ | Coupling, vector bulk space-time 4-form |
| $\Phi^i$ | $m^i = \partial \mathcal{L}_\Phi / \partial \Phi^i$ | Constitutive, axial vector bulk 3-form |
| $A_k^i$ | $J_k^i = \partial \mathcal{L}_Y / \partial A_k^i = \epsilon_{ilm} v_l \sigma_{km}$ | Coupling, axial vector space-time 3-form |
| $Y_{ki}$ | $\sigma_{ki} = \partial \mathcal{L}_Y / \partial Y_{ki}$ | Constitutive, vector area space-time 3-form |
| $\chi_{ki}$ | $\Pi_{ki} = \partial \mathcal{L}_\chi / \partial \chi_{ki}$ | Constitutive, vector space-time 2-form |
| $H_k^i$ | $E_k^i = \partial \mathcal{L}_H / \partial H_k^i$ | Constitutive, axial vector area 2-form |
| $B_k^i$ | $G_k^i = \partial \mathcal{L}_\chi / \partial B_k^i = \epsilon_{ilm} v_l \Pi_{km}$ | Coupling, axial vector space-time 2-form |

## 4. Dynamical equations of viscous flow

Use $\varphi_i, \eta_k^i$ and $\eta_4^i$ as the variations of $v_i, A_k^i$ and $\Phi^i$, respectively, namely

$$\varphi_i = \delta v_i, \eta^i_k = \delta A^i_k, \eta^i_4 = \delta\Phi^i,$$

so, from

$$\delta Y_{ki} = \partial_k \varphi_i + \epsilon_{ipq} A^p_k \varphi_q + \epsilon_{ipq} \eta^p_k v_q,$$
$$\delta B^i_k = \epsilon_{klm}\left(\partial_l \eta^i_m + \epsilon_{ipq} \eta^p_l A^q_m\right),$$
$$\delta H^i_k = \partial_k \eta^i_4 - \partial_t \eta^i_k + \epsilon_{ipq} \eta^p_l \Phi^q + \epsilon_{ipq} A^p_k \eta^q_4,$$

we have the variation of the Lagrangian density respect to $\{v_i; W^j_\mu\}$ as

$$\begin{aligned}
\delta\mathcal{L} = (M_i + \Sigma_i)\varphi_i + \sigma_{ki}\delta Y_{ki} - m^i \eta^i_4 + G^i_k \delta B^i_k - E^i_k \delta H^i_k \\
&= \left(M_i + \Sigma_i - \partial_k \sigma_{ki} - \epsilon_{ilm} A^l_k \sigma_{km}\right)\varphi_i + \partial_k\left(\sigma_{ki}\varphi_i\right) \\
&- \left(m^i - \partial_k E^i_k - \epsilon_{ipq} A^p_k E^q_k\right)\eta^i_4 - \partial_k\left(E^i_k \eta^i_4\right) \\
&+ \left[J^i_k - \partial_t E^i_k - \epsilon_{ipq}\Phi^p E^q_k + \epsilon_{klm}\left(\partial_l G^i_m + \epsilon_{ipq} A^p_l G^q_m\right)\right]\eta^i_k + \partial_t\left(E^i_k \eta^i_k\right) \\
&- \partial_k\left(\epsilon_{klm} G^i_l \eta^i_m\right).
\end{aligned}$$

Noticing that the variations of field variables are independent to each other and vanish on the boundary, the minimal action principle of fluid system in the space-time domain $\wp$ occupied by the fluid

$$\delta \int_\wp \mathcal{L}[v_i; A^i_k, \Phi^i] dv \wedge dt = \int_\wp \delta\mathcal{L}[v_i; A^i_k, \Phi^i] dv \wedge dt = 0 \tag{15}$$

finally yields the dynamical equations between the generalized forces, which can be compactly written with the differential forms as

$$\mathbf{M}_i + \mathbf{\Sigma}_i = d\boldsymbol{\sigma}_i + \epsilon_{ilm}\mathbf{A}^l \wedge \boldsymbol{\sigma}_m \triangleq D\boldsymbol{\sigma}_i, \tag{16}$$
$$\mathbf{J}^i = d\mathbf{Q}^i + \epsilon_{ipq}\mathbf{W}^p \wedge \mathbf{Q}^q \triangleq D\mathbf{Q}^i, \tag{17}$$

where

$$\mathbf{J}^i = \frac{\partial\mathcal{L}_\Phi}{\partial\Phi^i} dv + \frac{\partial\mathcal{L}_Y}{\partial A^i_k} da_k \wedge dt = \mathbf{m}^i + \epsilon_{ilm}\boldsymbol{\sigma}_l v_m, \tag{18.1}$$

$$\mathbf{Q}^i = \frac{\partial\mathcal{L}_H}{\partial H^i_k} da_k + \frac{\partial\mathcal{L}_\chi}{\partial B^i_k} dx_k \wedge dt = \mathbf{E}^i + \epsilon_{ilm}\mathbf{\Pi}_l v_m. \tag{18.2}$$

It should be pointed out that the exterior differentials (16) are confined to be of spatial, since the force equilibrium holds on any volume domain at every instant. But in (17), there are two parts: one is about the micro-moments of force over the volume domain, another is about the micro-moment fluxes of force across the surface domain. It is interesting that since the two parts of micro-moments of force are not completely independent, the derived integrability condition

$$D\mathbf{J}^i = d\mathbf{J}^i + \epsilon_{ipq}\mathbf{W}^p \wedge \mathbf{J}^q = \epsilon_{ipq}\mathbf{F}^p \wedge \mathbf{Q}^q = \epsilon_{ipq}\left(B^p_k G^q_k + H^p_k E^q_k\right) dv \wedge dt \tag{19}$$

is usually simplified to covariant conservation of the micro-moment current

$$D\mathbf{J}^i = 0 \tag{20}$$

if the coaxiality between **H** and **E**, **B** and **G** are satisfied, respectively. By virtue of coaxiality between **m** and **Φ**, **σ** and **Y**, further expansion of (20) yields

$$\partial_t m^i = \epsilon_{ipq} v_p D_k \sigma_{kq} \tag{21}$$

implying that the bulk twirling interaction directly produces the micro-eddies.

Different fluids may possess different constitutive laws and different properties of matter. For most fluids consisting of small molecules, the intrinsic isotropy will guarantee the coaxiality between the generalized flux and force in the constitutive law. Further for the Newton's fluids, like water and air in ordinary flows, the Lagrangian density is simply quadratic (or the constitute law is linear [10]) such that

$$\mathcal{L}[v_i; W_\mu^i] = \mathcal{L}_{Euler}[v_i] + \frac{1}{2}\mu\left(Y_{ki}Y_{ki} - \Phi^i\Phi^i\right) + \frac{1}{2}\mu\Lambda\left(\chi_{ki}\chi_{ki} - H_k^i H_k^i\right) \tag{22}$$

contains only two parameters of matter: the familiar dynamic viscosity $\mu$ and a new parameter $\Lambda$ with the dimension of area, which could be related to the characteristic size of micro-finite element. Under the construction (8) of Lagrangian density, we derive the controlling equations

$$\rho\frac{dv_i}{dt} - f_i + \partial_i p = \mu\nabla^2 v_i + \mu\epsilon_{ilm}\left(2A_k^l\partial_k v_m + v_m\partial_k A_k^l\right)$$

$$-\mu h_{ij,mn} v_j (A_k^m A_k^n + \Lambda B_k^m B_k^n), \tag{23}$$

$$\Phi^i = \Lambda\left(\partial_k H_k^i + \epsilon_{ipq} A_k^p H_k^q\right), \tag{24}$$

$$\epsilon_{ilm} v_l Y_{km} = \Lambda\left(\partial_t H_k^i + \epsilon_{ipq}\Phi^p H_k^q\right) - \Lambda\epsilon_{klm}\left[\partial_l\left(\epsilon_{ipq} v_p \chi_{mq}\right) + \epsilon_{irs}\epsilon_{spq} A_l^r v_p \chi_{mq}\right] \tag{25}$$

for the velocity $v_i$, micro-eddy $\Phi^i$ and swirl $A_k^i$, respectively, where the viscosity coefficient in (24) and (25) have been eliminated from both sides, and

$$h_{ij,mn} = \epsilon_{kim}\epsilon_{kjn} = \delta_{ij}\delta_{mn} - \delta_{in}\delta_{jm} \tag{26}$$

As mentioned above, these equations represent the balances of momentum and micro-moments in fluid. The equations (23), (24) and (25) together with the continuity equation $\partial_i v_i = 0$ constitute a close system of sixteen equations for the sixteen field quantities $\{p, v_i, A_k^i, \Phi^i, i, k = 1,2,3\}$. Finally, it is easy to verify that the conservation (21) of micro-moment current takes the form

$$\partial_t \Phi^i = \epsilon_{ipq} v_p D_k Y_{kq}, \tag{27}$$

seeming to be a geometrical relation without any property of matter.

# 5. Preliminary analyses of dynamical equations

## 5.1. Energy equilibriums

Rewrite the dynamical equations (16) and (17) in the component form, namely

$$\rho a_i - f_i + \partial_i p + \Sigma_i = \partial_k \sigma_{ki} + \epsilon_{ilm} A_k^l \sigma_{km}, \tag{28}$$

$$m^i = \partial_k E_k^i + \epsilon_{ilm} A_k^l E_k^m, \tag{29}$$

$$\epsilon_{ilm} v_l \sigma_{km} = \partial_t E_k^i + \epsilon_{ipq}\Phi^p E_k^q - \epsilon_{klm}\left(\partial_l G_m^i + \epsilon_{ipq} A_l^p G_m^q\right), \tag{30}$$

we can derive the energy equilibrium relations

$$\rho a_i v_i - f_i v_i + v_i \partial_i p + \Sigma_i v_i = \partial_k(\sigma_{ki} v_i) - \sigma_{ki} Y_{ki}, \tag{31}$$

$$m^i \Phi^i = \partial_k\left(E_k^i \Phi^i\right) - E_k^i \partial_t A_k^i - E_k^i H_k^i, \tag{32}$$

$$\epsilon_{ilm}A^i_k v_l \sigma_{km} = \partial_t\left(E^i_k A^i_k\right) - E^i_k \partial_k \Phi^i + E^i_k H^i_k - \epsilon_{klm}\partial_l\left(G^i_m A^i_k\right) - G^i_k B^i_k - \frac{1}{2}\epsilon_{klm}\epsilon_{ipq}A^p_k A^q_l G^i_m. \quad (33)$$

From the first two relations, we find the non-negative dissipative terms $\sigma_{ki}Y_{ki}$, $\Sigma_i v_i$ and $E^i_k H^i_k$, and the boundary acting terms $\partial_k(\sigma_{ki}v_i)$ and $\partial_k(E^i_k\Phi^i)$; it is interesting that there appears a term $E^i_k\partial_t A^i_k$ from the changing of the swirl field in (32). The third equation is of the energy transfer: two non-negative terms $E^i_k H^i_k$ and $G^i_k B^i_k$, having different signs here, indicate the exchange of energy between the swirl structures and the micro-eddies; the term $\partial_t\left(E^i_k A^i_k\right)$ can be regarded as the increase of energy stored in the swirl field while the term $\epsilon_{ilm}A^i_k v_l \sigma_{km}$ shows the contribution of the twirling process from the main stream. The combination of the three energy equations results in the equilibrium of total mechanical energy as

$$\frac{d\left(\frac{1}{2}\rho v_i v_i\right)}{dt} + \sigma_{ki}Y_{ki} + m^i\Phi^i + E^i_k H^i_k = f_i v_i + \partial_k\left(p v_k + \sigma_{ki}v_i + E^i_k\Phi^i\right) - \partial_t\left(E^i_k A^i_k\right)$$
$$+\epsilon_{klm}\partial_l\left(G^i_m A^i_k\right) + \epsilon_{ilm}A^i_k v_l \sigma_{km} + \epsilon_{ipq}\Phi^p A^q_k E^i_k + \frac{1}{2}\epsilon_{klm}\epsilon_{ipq}A^p_k A^q_l G^i_m, \quad (34)$$

where uses are made of the continuity equation and $\Sigma_i v_i = G^i_k B^i_k$. The increase of kinetic energy and the dissipative terms are listed in the left, whilst the right, besides the works of body force, pressure and boundary viscous forces, includes three terms which can be understood as the coupling works of different fields.

## 5.2. Field equations under Frenét frames of instant streamlines

A streamline at the instant $t$ and across the point $\mathbf{r}_0$ is a one-parameter space curve defined by

$$\frac{d\mathbf{r}(\tau;t)}{d\tau} = \mathbf{v}(\mathbf{r},t) = V\mathbf{n}, \mathbf{r}(0;t) = \mathbf{r}_0, \quad (35)$$

where $V \geq 0$ is the speed, and the unit vector $\mathbf{n}$ indicates the direction of flow. The streamline at a point with non-zero speed is unique, and independent of the speed. Except for few isotropic points with $V = 0$, the flow field around them requires special attention in the future, we focus on the regular domain with $V > 0$ and introduce $ds = Vd\tau$, then the streamline equation can be rewritten as

$$\frac{d\mathbf{r}(s;t)}{ds} = \mathbf{n}(\mathbf{r},t), \mathbf{r}(0;t) = \mathbf{r}_0, \quad (36)$$

where the one-parameter becomes the arc length of the streamline starting from $\mathbf{r}_0$.

According to the differential geometry, the characteristic of the streamline as a space curve can be determined by the arc length $s$, the curvature $\kappa$ and torsion $\omega$ (if the curvature is non-zero) within a rigid motion (a translation plus a rotation). Denote by

$$\partial_n = \mathbf{n}\cdot\nabla, \quad (37)$$

we have the relations

$$\boldsymbol{\xi}_1 = \mathbf{n};\ \partial_n\begin{pmatrix}\boldsymbol{\xi}_1\\ \boldsymbol{\xi}_2\\ \boldsymbol{\xi}_3\end{pmatrix} = \begin{pmatrix}0 & \kappa & 0\\ -\kappa & 0 & \lambda\\ 0 & -\lambda & 0\end{pmatrix}\begin{pmatrix}\boldsymbol{\xi}_1\\ \boldsymbol{\xi}_2\\ \boldsymbol{\xi}_3\end{pmatrix}. \quad (38)$$

under the Frenét (orthogonal) frame $\{\boldsymbol{\xi}_1, \boldsymbol{\xi}_2, \boldsymbol{\xi}_3\}$. In the regular domain, the streamlines are uniquely determined when the contact or statistical velocity field is given.

The slip laminarization of fluid originates from the nonuniform flow and couples with the velocity. Considering that the fluid cannot resist shear, and the material density of the fluid in the incompressible flow is constant in the macroscopic statistic, the ordering of the fluid can only be the slip ordering, such that the general linear transformation group expressing the local isomorphism of the minimum units of flow analysis, namely the micro-finite elements, must be the rotation group. Therefore, the elements are no longer assumed to be isolated fluid particles, but open sets that can overlap each other (Fig. 2). A fluid element consists of a cluster of fluid molecules with orientation characteristics, which are definite in local space-time. However, the orientation of fluid elements itself has no absolute significance, but the orientation difference of adjacent fluid elements defines the isomorphic translation of fluid elements. In laminar flows, where fluid is steadily transported along the streamlines, we have $\Phi^i = H_k^i = B_k^i = 0$, and $\epsilon_{ilm} v_l Y_{km} = 0$. The last constraint results in the swirl

$$A_k^i = -\epsilon_{ilm} n_l \partial_k n_m$$

within any term satisfying $n_i A_k^i \neq 0$, which means that the swirl components around the direction of velocity doesn't affect the coupling between the vortex and the mainstream.

It is interesting to investigate the fluid dynamics under the Frenét frame $\{\boldsymbol{\xi}_1, \boldsymbol{\xi}_2, \boldsymbol{\xi}_3\}$. Assume that

$$\boldsymbol{\xi}_i = R_{ij} \mathbf{e}_j, \tag{39}$$

using the definition

$$D\boldsymbol{\xi}_i = \epsilon_{ipq} \widetilde{\mathbf{W}}^p \boldsymbol{\xi}_k, \tag{40}$$

we can derive the connection and the curvature tensor under the Frenét frames as

$$\widetilde{\mathbf{W}}^i = R_{ip} \mathbf{W}^p - \mathbf{w}^i, \widetilde{\mathbf{F}}^i = D\widetilde{\mathbf{W}}^i = R_{ip} \mathbf{F}^p, \tag{41}$$

with

$$\mathbf{w}^i = \epsilon_{ipq} \frac{1}{2} R_{pk} dR_{qk} \leftrightarrow \epsilon_{ipq} \mathbf{w}^i = R_{pk} dR_{qk}. \tag{42}$$

The transformation relations (41) show that the curvature tensor is really a tensor covariantly changing with the frame while the connection is not. It is remarkable that all terms after covariant differential are covariant, for instance we have the transformations of the generalized fluxes $\mathbf{Y}_i$ and $\boldsymbol{\chi}_i$ as

$$\widetilde{\mathbf{Y}}_i = R_{ip} \mathbf{Y}_p, \tilde{\boldsymbol{\chi}}_i = R_{ip} \boldsymbol{\chi}_p. \tag{43}$$

As for the quantities $\mathbf{M}_i$ and $\mathbf{m}^i$, there exist a kinematic explanation to guarantee their covariance, namely the missing acceleration and micro-eddy observed in the space-time inhomogeneous frame must be picked up in the dynamical equilibriums. Therefore, the dynamical equations possess the *form invariance* as

$$\rho(\tilde{a}_i + \epsilon_{il1} \Omega^l V) - \tilde{f}_i + \tilde{\partial}_i p + \tilde{\Sigma}_i = \partial_k \tilde{\sigma}_{ki} + \epsilon_{ilm} \tilde{A}_k^l \tilde{\sigma}_{km}, \tag{44}$$

$$\widetilde{m}^i = \partial_k \tilde{E}_k^i + \epsilon_{ilm} \tilde{A}_k^l \tilde{E}_k^m, \tag{45}$$

$$\epsilon_{i1m} V \tilde{\sigma}_{km} = \partial_t \tilde{E}_k^i + \epsilon_{ipq} \tilde{\Phi}^p \tilde{E}_k^q - \epsilon_{klm} \left( \partial_l \tilde{G}_m^i + \epsilon_{ipq} \tilde{A}_l^p \tilde{G}_m^q \right), \tag{46}$$

where

$$\Omega^i = -w_4^i - v_k w_k^i \tag{47}$$

coming from the relation

$$\tilde{a}_i + \epsilon_{il1} \Omega^l V = R_{ip} \frac{dv_p}{dt} = \delta_{i1} \frac{dV}{dt} + V R_{ip} \frac{dR_{1p}}{dt} \tag{48}$$

For the Newton's fluids, the field equations become

$$\rho \left( \delta_{i1} \frac{dV}{dt} + \epsilon_{il1} \Omega^l V \right) - \tilde{f}_i + \tilde{\partial}_i p = \mu \delta_{i1} \nabla^2 V + \mu \epsilon_{il1} \left( 2 \tilde{A}_k^l \partial_k V + V \partial_k \tilde{A}_k^l \right)$$

$$-\mu h_{i1,mn} V \left( \tilde{A}_k^m \tilde{A}_k^n + \Lambda \tilde{B}_k^m \tilde{B}_k^n \right), \tag{49}$$

$$\tilde{\Phi}^i + w_4^i = \Lambda \left( \partial_k \tilde{H}_k^i + \epsilon_{ilm} \tilde{A}_k^l \tilde{H}_k^m \right), \tag{50}$$

$$V^2 \left( \tilde{A}_k^i - \delta_{i1} \tilde{A}_k^1 \right) = \Lambda \left( \partial_t \tilde{H}_k^i + \epsilon_{ipq} \tilde{\Phi}^p \tilde{H}_k^q \right)$$

$$-\Lambda \epsilon_{klm} \left\{ \partial_l \left[ V^2 \left( \tilde{B}_m^i - \delta_{i1} \tilde{B}_m^1 \right) \right] + \epsilon_{ipq} \tilde{A}_l^p V^2 \left( \tilde{B}_m^q - \delta_{q1} \tilde{B}_m^1 \right) \right\}, \tag{51}$$

The conservation relation (27) under the Frenét frame becomes

$$\partial_t \left( \tilde{\Phi}^i + w_4^i \right) + \epsilon_{ipq} \tilde{\Phi}^p w_4^q = \partial_k \left[ V^2 \left( \tilde{A}_k^i - \delta_{i1} \tilde{A}_k^1 \right) \right] + \epsilon_{i1p} V^2 \tilde{A}_k^1 \tilde{A}_k^p, \tag{52}$$

The streamwise projections of equations (49), (51) and (52)

$$\rho \frac{dV}{dt} - \tilde{f}_1 + \tilde{\partial}_1 p = \mu \nabla^2 V - \mu \left( \tilde{A}_k^2 \tilde{A}_k^2 + \tilde{A}_k^3 \tilde{A}_k^3 + \Lambda \tilde{B}_k^2 \tilde{B}_k^2 + \Lambda \tilde{B}_k^3 \tilde{B}_k^3 \right) V, \tag{53}$$

$$0 = \partial_t \tilde{H}_k^1 + \tilde{\Phi}^2 \tilde{H}_k^3 - \tilde{\Phi}^3 \tilde{H}_k^2 - \epsilon_{klm} V^2 \left( \tilde{A}_l^2 \tilde{B}_m^3 - \tilde{A}_l^3 \tilde{B}_m^2 \right), \tag{54}$$

$$\partial_t \left( \tilde{\Phi}^1 + w_4^1 \right) + \tilde{\Phi}^2 w_4^3 - \tilde{\Phi}^3 w_4^2 = 0 \tag{55}$$

reveal the following information: (1) the flow has a resistance proportional to the speed if there are swirl structures perpendicular to the flow, (2) the origin of the streamwise vortex fields are completely geometrical, namely from the non-interchangeability of rotation group. Their projections perpendicular to the stream can be written by ($i \neq 1$)

$$\rho \Omega^i V + \epsilon_{i1m} \left( \tilde{\partial}_m p - \tilde{f}_m \right) = \mu V^{-1} \partial_k \left( V^2 \tilde{A}_k^i \right) + \mu \epsilon_{i1m} V \left( \tilde{A}_k^m \tilde{A}_k^1 + \Lambda \tilde{B}_k^m \tilde{B}_k^1 \right), \tag{56}$$

$$V^2 \tilde{A}_k^i = \Lambda \left( \partial_t \tilde{H}_k^i + \epsilon_{ipq} \tilde{\Phi}^p \tilde{H}_k^q \right) - \Lambda \epsilon_{klm} \left[ \partial_l \left( V^2 \tilde{B}_m^i \right) + \epsilon_{ipq} \tilde{A}_l^p V^2 \left( \tilde{B}_m^q - \delta_{q1} \tilde{B}_m^1 \right) \right], \tag{57}$$

$$\partial_t \left( \tilde{\Phi}^i + w_4^i \right) + \epsilon_{ipq} \tilde{\Phi}^p w_4^q = \partial_k \left( V^2 \tilde{A}_k^i \right) + \epsilon_{i1p} V^2 \tilde{A}_k^1 \tilde{A}_k^p \tag{58}$$

The combination of (56) and (57)

$$\mu \partial_t \left( \tilde{\Phi}^i + w_4^i \right) + \mu \epsilon_{ipq} \tilde{\Phi}^p w_4^q = V \left[ \rho \Omega^i V + \epsilon_{i1m} \left( \tilde{\partial}_m p - \tilde{f}_m \right) \right] - \mu \Lambda \epsilon_{i1m} V^2 \tilde{B}_k^m \tilde{B}_k^1 \tag{59}$$

shows that the generation of non-streamwise micro-eddies may come from the non-parallelism between the resultant force of non-viscous force and stream, or the coupling of the swirl intensity along different directions.

## 5.3. Expansions of field equations according to master variables and their one-dimensional models

All field equations (23)-(25) can be written in a four-term standard form according to their master variables as

$$\textbf{Unsteady term} + \textbf{Linear term} + \textbf{Diffusion term} = \textbf{Source term}\,. \tag{1}$$

In order to express the expansion in a compact form as possible, we introduce the following notations: (i) maintain all terms in the velocity equations if they are the same as in the classical theory; (ii) keep their original expression if they obviously include no master variable, (iii) for a free index $i$, use $(i, i_1, i_2)$ as an even permutation of $(1,2,3)$, for example $(i, i_1, i_2)$ corresponds to $(2, 3, 1)$ if $i = 2$, the sum for repeated indices $i_s$ is from 1 to 2, (iv) use $\langle i_1, i_2 \rangle$ for the sum in an asymmetrical way, for example

$$\Phi^{\langle i_1}\partial_t A_k^{i_2\rangle} = \Phi^{i_1}\partial_t A_k^{i_2} - \Phi^{i_2}\partial_t A_k^{i_1}, \tag{61}$$

and (v) for the repeated indices with underscore the sum convention no longer works. Then, after some lengthy arrangement, we obtain the following equations

$$\rho\frac{dv_i}{dt} + \mu\left(A_k^{i_s}A_k^{i_s} + \Lambda B_k^{i_s}B_k^{i_s}\right)v_i - \mu\nabla^2 v_i = f_i - \partial_i p + \mu\epsilon_{ilm}\left(2A_k^l\partial_k v_m + v_m\partial_k A_k^l\right)$$
$$+\mu\left(A_k^i A_k^{i_s} + \Lambda B_k^i B_k^{i_s}\right)v_{i_s}, \tag{62}$$

$$\left(\Lambda^{-1} + A_k^{i_s}A_k^{i_s}\right)\Phi^i - \nabla^2\Phi^i = A_k^i A_k^{i_s}\Phi^{i_s} - \partial_t\partial_k A_k^i + \epsilon_{ilm}\left[A_k^l(2\partial_k\Phi^m - \partial_t A_k^m) - \Phi^m\partial_k A_k^m\right], \tag{63}$$

$$\partial_t\partial_t A_k^i + \left[\Lambda^{-1}v_{i_s}v_{i_s} + v_{\underline{i}}v_{\underline{i}}A_{k_t}^{i_s}A_{k_t}^{i_s} - \Phi^{i_s}\Phi^{i_s} + v_{i_s}v_{i_r}A_{k_t}^{i_s}A_{k_t}^{i_r} + \partial_{k_t}\left(v_i v_{\langle i_1}A_{k_t}^{i_2\rangle}\right)\right]A_k^i$$
$$-\partial_{k_t}\left(v_{i_s}v_{i_s}\partial_{k_t}A_k^i\right) = \Lambda^{-1}\left(v_i v_{i_s}A_k^{i_s} - v_{\langle i_1}\partial_k v_{i_2\rangle}\right) + \partial_t\left(\partial_k\Phi^i + A_k^{\langle i_1}\Phi^{i_2\rangle}\right) - \Phi^i\Phi^{i_s}A_k^{i_s}$$
$$+\Phi^{\langle i_1}\left(\partial_k\Phi^{i_2\rangle} - \partial_t A_k^{i_2\rangle}\right) - \partial_{k_t}\left[v_{i_s}v_{i_s}\left(\partial_k A_{k_t}^i + A_k^{\langle i_1}A_{k_t}^{i_2\rangle}\right) - v_i v_{i_s}\partial_{\langle k}A_{k_t\rangle}^{i_s} - v_{\langle i_1}A_k^{i_2\rangle}v_p A_{k_t}^p\right]$$
$$-V^2 A_{k_t}^{\langle i_1}\partial_{\langle k}A_{k_t\rangle}^{i_2\rangle} + A_{k_t}^{\langle i_1}v_{i_2\rangle}v_p\partial_k A_{k_t}^p + \left(v_{\underline{i}}v_{\underline{i}}A_{k_t}^{i_s}A_k^{i_s} + v_{i_s}v_{i_r}A_{k_t}^{i_s}A_k^{i_r}\right)A_{k_t}^i + A_{k_t}^{\langle i_1}v_{i_2\rangle}A_{\langle k}^{i_1}A_{k_t\rangle}^{i_2}v_i, \tag{64}$$

for the principal governing properties $\{v_i, \Phi^i, A_k^i\}$, where the complicated coefficients and sources show all coupling between multi-fields and multi-components.

In view of the principal variables, there is no non-linear coupling of fields themselves, since the conventional coupling in the convection term of the velocity can be ascribed to the pressure, say in the derivation

$$\rho\frac{dv_i}{dt} + \partial_i p = \rho\partial_t v_i + \rho v_{i_s}\left(\partial_{i_s}v_i - \partial_i v_{i_s}\right) + \partial_i\left(p + \frac{1}{2}\rho V^2\right), \tag{65}$$

the second term in the right has no coupling of the principal variable with itself. Other obvious features of the governing equations include: (1) the equations the swirl fields have two-order unsteady term while the equations of the micro-eddy fields have no unsteady term, (2) the swirl fields diffuse transversely, (3) the coefficients of linear terms for the fields associated with motion (velocity and micro-eddy) are non-negative while those for the fields associated with structure (swirl) are indefinite. The existence of linear term is the key point of new theory. The linear term could be a regulator between the temporal evolution and spatial distribution: a positive coefficient of linear term means exponential attenuation with time and localization in space while a negative one proclaims wavy change in space and with time if the time derivative is two order, or exponential increase with time if the time derivative is one order.

It is interesting to truncate the field equations to yield their one-dimensional models. We find that there are three types of models

$$\partial_t V + \nu k^2 V - \nu\partial_{xx}V = S_V, \tag{66}$$

$$(a^{-2} + k^2)\Phi - \nu\partial_{xx}\Phi = S_\Phi, \tag{67}$$

$$c^{-2}\partial_{tt}A \pm k^2A - \partial_{xx}A = S_A, \tag{68}$$

for the velocity, micro-eddy and swirl fields, respectively, where $\nu$ and $a$ are properties of the fluid, $k$ and $c$ are parameters depending on the flow. These model equations can be used to discuss the evolution characteristic of different fields.

# 6. Expectable experiment verifications

For a parallel flow with linear shear, say the Couette flow between two plates (Fig. 3), the theory of the N-S equations admits that two pairs of shear stresses equal to each other on the surfaces make the element distort and rotate, as shown in Fig. 4(bottom), and so decrease in height in the infinitesimal deformation analysis [11]. How strange is such a model? The fluid particle cannot be an element with finite size: the continuous distortion makes it untraceable. The decrease in height conflicts with our intuition and observation. The theory presented in this people tells a different story about this simple flow (Fig. 4(top)). The fluid element slips under the shear friction, and cannot be regarded as a standard fluid element after any finite time, since any fluid element is defined in a local space-time and the fluid elements on different positions can be assigned independently.

Many differences could be found between two theories when the flow being curved, unsteady or even turbulent. Some polymer additives will change their molecular configurations under the shear circumstance of flow [12][13], which can be made use of to visualize the structures of viscous interaction. For example, two flows shown in Fig. 1 are the same in their inner shear interactions according to the classical theory of flow, can we visualize them to confirm the judgement? Here we propose an experiment which is expected to help checking the rationality of the new theory of flow.

According the slip mechanism of viscosity exploited in this paper, no bulk of fluid can sustain a rigid motion only through the viscous interaction. Let us consider a viscous fluid confined in the gap between two coaxial cylinders. According to the theory of the N-S equations, the fluid will move like a rigid body when two cylinders rotate with the same angular velocity. But under the new theory, the speed of flow is given by

$$V = \Omega R_1 + \Omega(R_2 - R_1)\frac{\ln r - \ln R_1}{\ln R_2 - \ln R_1} \neq \Omega r. \tag{69}$$

So, the precise measurement of the velocity profile will provide a good verification for the new theory. In practice, the length of the cylinder cannot be infinite, and the requirement for the stability of flow also needs a limit on the scale of gap and the angular velocity. van Gils *et al.* [14][15] tested the predicted rigid motion of water between two cylinders with height of $0.927\mathrm{m}$ under the parameters

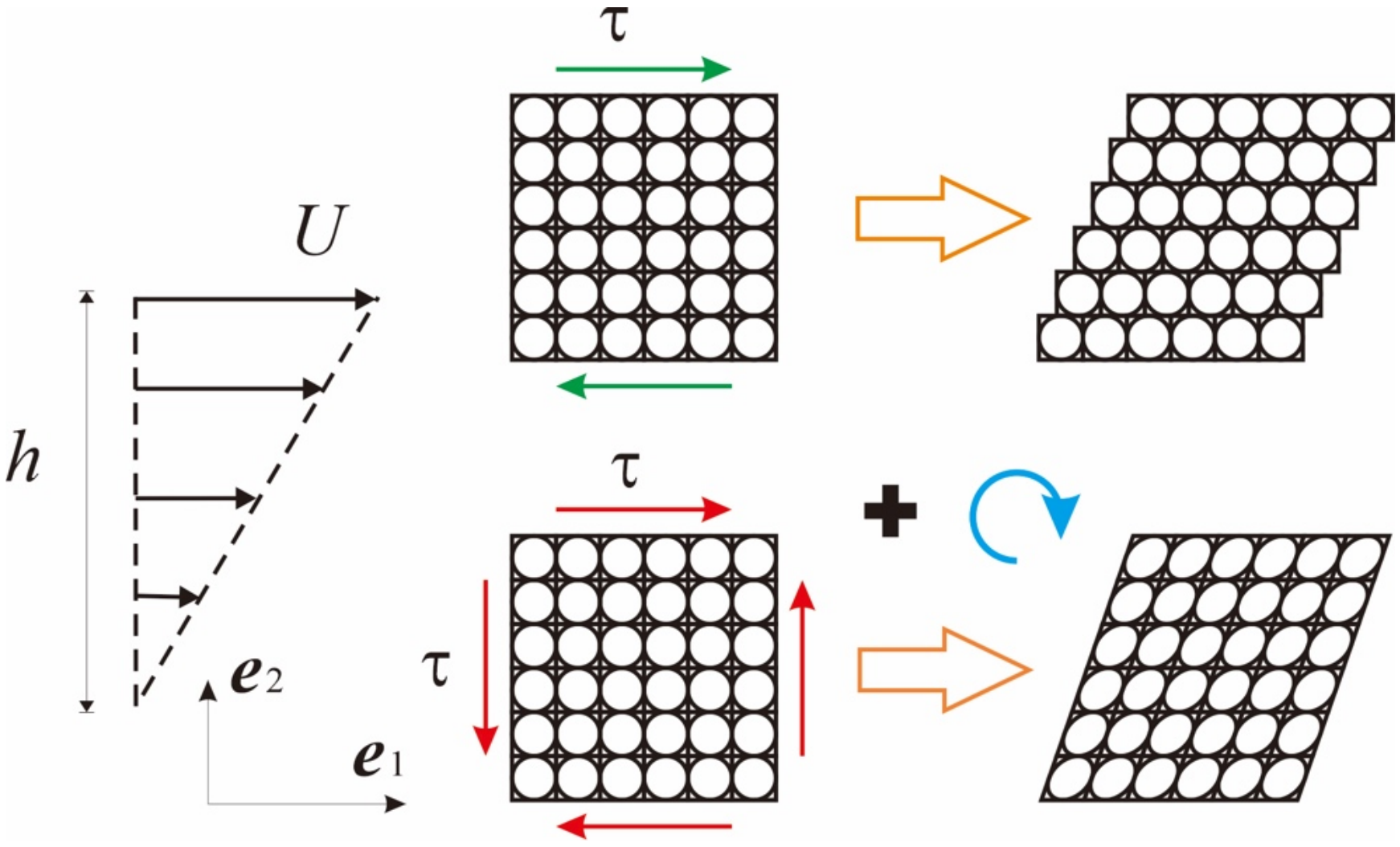

Fig. 4. Different understandings of the simple shear flow.

$$R_1 = 0.2\text{m}, R_2 = 0.279\text{m}, \Omega = 2\text{Hz}. \tag{70}$$

They measured the azimuthal velocity profile with LDA and found some errors within 0.6% from the rigid motion. Using the formula (69), we can derive the maximal errors from the rigid motion are from 0.11% to 1.42% when the ratio $R_2/R_1$ changes from 1.1 to 1.4. The error increases quickly with the radius ratio, say it reaches 4% if the ratio is 1.8, which is quite obvious to find using today's technology.

## 7. Concluding remarks

When the fluid element occupying a space-time point and endowed with the macroscopic velocity is modeled to be of micro-finite instead of infinitesimal, whatever small it is, the description of flowing as a space-time manifold becomes natural and the dynamics of irreversibly topological evolution of contact structures in the fluid is constructed by expressing the viscous interaction as internal slip friction. In the new framework, the coherent structures can be recognized through the nontrivial (non-integrable) swirl fields, and the domain of turbulence can be defined as the fluid with small-scale eddy. The illusions of vortex pattern observed by the moving observers [16] will be clarified since the ordering structures couple with the velocity field that is uniquely determined. Therefore, the spatial pattern and the temporal dissipation are perfectly combined to a unified quantity as the space-time connection between the fluid elements.

In summary, a new theory of viscous flow is elaborated based on the micro-finite elements and the slip and twirling mechanism of viscous interactions. Unlike the corresponding theory of elasticity, here is no topologically equivalence between theory of micro-finite element and that of particle. The new variables, namely the vortex fields characterizing the structures between elements and the micro-eddies in the element, make the flowing fluid a space-time differential manifold covered by the micro-finite elements. Under the new description of fluid flow, the main stream, vortex structures and micro-eddies constitute a strongly coupled system, where the nonlinear couplings between the components of different directions, some of them are completely geometrical, take a critical role. Under the assumption of linear constitutive laws, only a new property of matter, with the measure of scale square, is needed to make the controlling field equations close. Some features of fluid flow are exploited through the analyses of the derived equations. Finally, an experiment is proposed to verify the rationality of new theory.

## Acknowledgments

Z.W.N. acknowledges the financial support from the NSFC (Grants No. 11372124).